\documentclass[lettersize,journal, twoside]{IEEEtran}
\usepackage{amsmath,amsfonts}
\usepackage{algorithmic}
\usepackage{algorithm}
\usepackage{array}
\usepackage[caption=false,font=normalsize,labelfont=sf,textfont=sf]{subfig}
\usepackage{textcomp}
\usepackage{stfloats}
\usepackage{url}
\usepackage{verbatim}
\usepackage{graphicx}
\usepackage{cite}
\usepackage[xindy]{glossaries}
\usepackage{siunitx}
\newacronym{dem}{DEM}{digital elevation model}
\newacronym{doa}{DoA}{direction-of-arrival}
\newacronym{fbp}{FBP}{Fast Back-Projection}
\newacronym{fft}{FFT}{Fast Fourier Transform}
\newacronym{fmcw}{FMCW}{Frequency-Modulated Continuous-Wave}
\newacronym{gps}{GPS}{Global Positioning System}
\newacronym{insar}{InSAR}{Interferometric Synthetic Aperture Radar}
\newacronym{mmwave}{mmWave}{millimeter wave}
\newacronym{pri}{PRI}{pulse repetition interval}
\newacronym{rx}{RX}{receive antenna}
\newacronym{tx}{TX}{transmit antenna}
\newacronym{vx}{VX}{virtual antenna}
\newacronym{rcs}{RCS}{Radar Cross Section}
\newacronym{sar}{SAR}{Synthetic Aperture Radar}
\newacronym{snr}{S/N}{signal-to-noise ratio}
\newacronym{tdm}{TDM MIMO}{Time-division multiplexing MIMO}
\newacronym{fov}{FOV}{field of view}

\usepackage{xcolor}

\usepackage[hidelinks]{hyperref}
\usepackage{orcidlink}

\begin{document}

\title{Automotive Elevation Mapping with Interferometric Synthetic Aperture Radar}

\author{Leyla A. Kabuli, Griffin Foster 
\thanks{The work of L. A. Kabuli was supported by the National Science Foundation's Graduate Research Fellowship under Grant DGE 2146752.}
\thanks{
L. A. Kabuli is with the Department of Electrical Engineering and Computer Sciences, University of California, Berkeley, CA 94720, USA (e-mail: lakabuli@berkeley.edu).
}
\thanks{
G. Foster is with Zendar Inc., Berkeley, CA 94710, USA (e-mail: gfoster@zendar.io).
}

} 

\maketitle
\thispagestyle{empty}
\markboth{Kabuli \MakeLowercase{et al.}}{Automotive Elevation Mapping with Interferometric Synthetic Aperture Radar}

\begin{abstract}
Radar is a low-cost and ubiquitous automotive sensor, but is limited by array resolution and sensitivity when performing direction of arrival analysis.
Synthetic Aperture Radar (SAR) is a class of techniques to improve azimuth resolution and sensitivity for radar. Interferometric SAR (InSAR) can be used to extract elevation from the variations in phase measurements in SAR images.
Utilizing InSAR we show that a typical, low-resolution radar array mounted on a vehicle can be used to accurately localize detections in 3D space for both urban and agricultural environments. 
We generate point clouds in each environment by combining InSAR with a signal processing scheme tailored to automotive driving.
This low-compute approach allows radar to be used as a primary sensor to map fine details in complex driving environments, and be used to make autonomous perception decisions.
\end{abstract}


\section{Introduction}

\IEEEPARstart{A}{utonomous} driving requires detailed mapping of the surroundings of a vehicle to infer drivable paths. Radar, lidar, camera, and ultrasonics are common sensors used for mapping. Radar arrays are low-cost sensors that provide high accuracy in range and Doppler of targets within the field of interest. However, lidar and camera typically provide higher angular resolution and accuracy. In radar, the target \gls{doa}, the angle estimation in azimuth and elevation of the target, is typically limited due to the physical array size and number of available elements. This limitation is a significant barrier to the utilization of radar for automotive use cases, and an active area of research~\cite{9369027}. Automotive applications require measuring information in the plane of vehicle motion. As a result, most automotive radar sensors prioritize azimuth over elevation in array design. However, for safe driving, elevation information is necessary. In any driving environment, there are overhanging structures (e.g. bridges, lights, trees, tunnels) and passable on-ground objects (e.g. lane markers, rocks, sewer covers, soil). Without elevation, the \gls{doa} does not have sufficient information to determine if it is possible to safely drive under, through, or over a detected object.

We present a new application of \gls{insar} combined with a real-time processing scheme that handles the intricacies of automotive driving. This approach provides enhanced sensitivity and angular accuracy in both azimuth and elevation of the environment. Building on our prior work~\cite{my_cisa_insar}, we study system accuracy and show results for general urban and agricultural environments, mapping a 3D world with human-scale features for use in safe autonomous driving. To our knowledge, this is the first work to do automotive \gls{insar} for elevation mapping of automotive environments.

The development of low-cost \gls{mmwave} radar sensors has enabled innovations in automotive radar approaches~\cite{9318758}.
One such approach is \gls{sar}, which forms high dynamic range images with sub-degree resolution across much of the azimuthal \gls{fov}. With a single \gls{vx}, formed from a \gls{tx} and \gls{rx}, azimuthal resolution is improved by utilizing the direction of travel to form a large, synthetic aperture~\cite{sar_chapter}. As the direction of travel is primarily on a plane, \gls{sar} does not provide any additional elevation resolution.

Interferometry is a technique that localizes a source by comparing the phase of signals received at two sensors. This pair of sensors, or baseline, measures a phase delay that depends on the source's location and the sensors' relative position and orientation. This phase delay can be used to determine the source's elevation. This method is commonly used in fields including optical imaging and radio astronomy~\cite{white_book, 2017isra.book.....T}. Interferometry is often performed using an array of sensors.

\gls{insar} combines \gls{sar} and interferometry~\cite{OGinsar} to form a 3D map of an environment. Interferometry can be performed using the \gls{vx} pairs on a radar array, forming baselines that further resolve a scene and provide elevation information. The choice of array layout with respect to the motion of travel is an important consideration when performing \gls{insar}. At typical driving speeds for automotive radar, the azimuth resolution is dominated by the \gls{sar} point response rather than the array point response in all directions except the extreme cone centered on the direction of travel~\cite{sar_chapter}. Additional baselines in the azimuth direction do not provide much in terms of resolution, but do provide additional sensitivity which is correlated with \gls{doa} accuracy. There is a significant advantage to using interferometry with elevation baselines as \gls{sar} alone does not provide enhanced resolution in that direction in the automotive case.

\gls{insar} is typically used to build \glspl{dem} of large regions and surfaces to measure the gradual evolution of geological features~\cite{groundterrain, terrainmap, unstableslopes}. As these imaging targets involve large physical and temporal scales, \gls{insar} is predominantly conducted at low frequency using airborne systems.

\gls{insar} is also incorporated in ground-based automotive radar systems, which offer more flexibility than rail-based systems~\cite{terrainmap}. A ground-based \gls{sar} system was mounted on a vehicle to map natural environments~\cite{korean_car_insar}. A car-borne \gls{sar} system was used to determine the slope of hills~\cite{simple_car_insar}. In contrast with these works, which focus on extracting elevation information for large, geological features, we map dense and detailed environments at close range, including agricultural and urban scenes. Additionally, we extract small, human-scale features for use in safe autonomous driving. Beamforming techniques have been used for 3D \gls{sar} imaging with elevation resolution enhancement using a vertical radar array~\cite{onlysarelevation}. Our elevation mapping approach is interferometric and utilizes an off-the-shelf radar array with minimal elevation components. While these works capture the environment on one side of the vehicle in an acquisition, our \gls{sar} system images the full frontal \gls{fov} of the vehicle in a single drive path. We use a robust and compact automotive \gls{sar} setup that is integrated with the vehicle.

In Section \ref{sec:sigmdl} we describe our signal model. In Section \ref{sec:setup} we describe our experimental setup and processing pipeline. In Section \ref{sec:results} we analyze the results of our driving scenarios and in Section \ref{sec:conclusion} discuss improvements and further work.

\section{Signal Model}
\label{sec:sigmdl}

In order to extract elevation, we start by forming \gls{sar} images from raw data. A separate \gls{sar} image is formed for each \gls{vx} on the real array, with a common phase center defined at the origin in order to maintain relative phase between \glspl{vx} at each coordinate position. Interferometric techniques are then used to extract phase delays from baselines on the real array, utilizing the complex-valued pixels in the \gls{sar} images. Source information, including elevation, is encoded in the phase delay for a baseline because the path length and resulting relative phase between a source and each \gls{vx} is different.

For our signal model, we assume a narrow fractional bandwidth of a few percent centered in the standard automotive radar band of 78 -- 81 \si{\giga\hertz} ($\lambda \approx 3.75$ \si{\milli\meter}).  The chirp used for imaging has a small fractional bandwidth such that we make the approximation that phase delays have no frequency dependence. We model sources as plane waves with a far-field approximation, as typical sources imaged are sufficiently far from the radar array when compared to the length of a baseline.

We assume that each range-azimuth bin in the \gls{sar} image contains at most one source. We can verify this assumption by considering the azimuth resolution of a \gls{sar} image, $\Psi_{SAR} = \frac{\lambda}{L \sin{\theta}}$ where $\lambda$ is the operational wavelength, $L$ is the aperture in meters, and $\sin{\theta}$ is the target azimuth angle. For a \SI{1}{\meter} aperture (reasonable for typical driving velocities) the azimuth resolution over much of the \gls{fov} is $< 0.25^{\circ}$ \cite{sar_chapter}. When combined with the typical range resolution on automotive radar (10--30 \si{\centi\meter}), the resulting \gls{sar} image bins are sufficiently small for most sources in automotive scenes. In the case of multiple sources within one bin, we fit the geometric mean of these sources. This fit will still be valid as long as there is one dominant source. In Section~\ref{sec:results}, our experimental results demonstrate that the fully sampled \gls{sar} image bins are sufficiently small for a one-source approximation to hold in many dense scenes. Future work could extend our approach to a multi-source model.

Under the signal model assumptions described above, the signal at each pixel in the \gls{sar} image is represented as a single complex-valued plane wave. For the \gls{sar} image formed for \gls{vx}\textsubscript{i}, the signal can be represented in terms of magnitude and phase as $S_i = |S_i|e^{j \psi_i}$.
Given signals $S_0$ and $S_1$ from the baseline formed with \gls{vx}\textsubscript{0} and \gls{vx}\textsubscript{1}, the phase delay at a pixel can be extracted from the complex correlation of the two signals
\begin{equation}
\label{eq:phasedelay}
  \Delta \psi = \angle (S_0 \cdot S_1^*)  = \psi_0 - \psi_1 .
\end{equation}

\subsection{Vertical Baseline Model} \label{vertical_model}
\begin{figure}[!t]
\centering 
\includegraphics[width=0.4\textwidth]{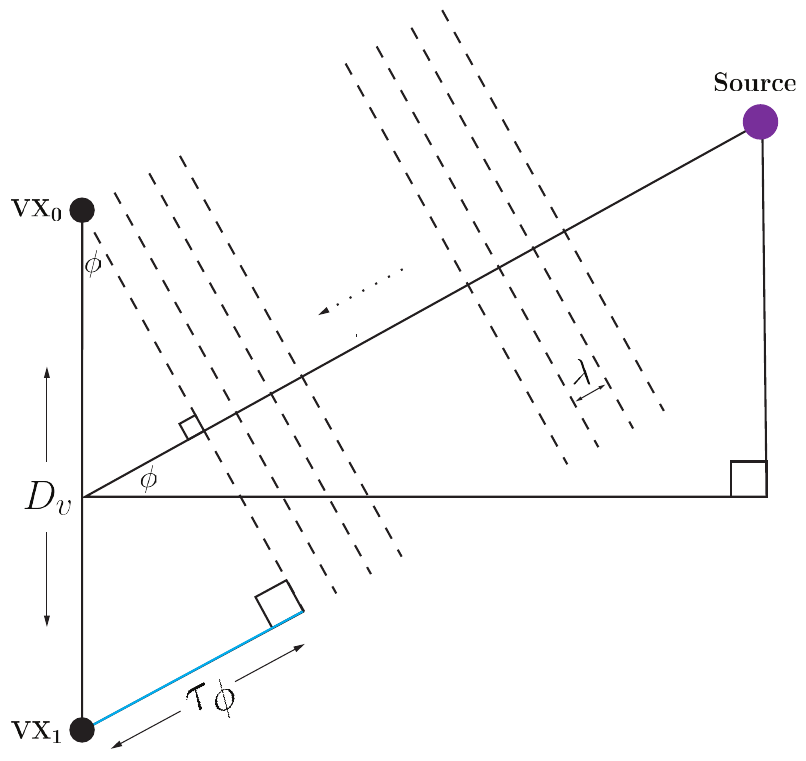}
\caption{Vertical Baseline Geometry: Illustration of the signal receive path from the source to the vertical baseline on the left. Dashed lines indicate the wavefront at regular intervals along the plane wave trajectory. The plane wave arrives at $\text{VX}_1$ with time delay $\tau_{\phi}$ relative to $\text{VX}_0$.}
\label{vertical_baseline_geometry}
\end{figure}

The interferometric model used to extract elevation information depends on the orientation of the baseline. The vertical baseline model is formed from two \glspl{vx} at the same horizontal offset and different elevation offsets in the real array. As shown in Figure~\ref{vertical_baseline_geometry}, the elevation offset between \glspl{vx} results in a time delay $\tau_{\phi}$, which depends on the baseline geometry and target elevation angle $\phi$. Using the vertical antenna spacing $D_v$ and the speed of light $c$, time delay and elevation angle are related by
\begin{equation}
  \tau_{\phi} = \frac{D_v}{c}\sin{\phi}. 
 \label{eq:tau_phi}
\end{equation}
As the signal received from a source is a phase delay, converting time delay to phase delay in Eq.~\eqref{eq:tau_phi} gives
\begin{equation}
\label{eq:phase_time_relation}
    \Delta{\psi} = 2 \pi \frac{c}{\lambda} \tau_{\phi} = 4 \pi \frac{D_v}{\lambda} \sin{\phi}.
\end{equation}
The additional factor of 2 accounts for the two directions in the transmit-receive path of the signals. The elevation angle is recovered from phase delay by solving for $\phi$ in Eq.~\eqref{eq:phase_time_relation}
\begin{equation}
\label{eq:phi_elev}
    \phi = \sin^{-1} \frac{\lambda}{4 \pi D_v} \Delta \psi.
\end{equation}
For an array that is dense in elevation, such that baselines are of $D_v \leq \frac{\lambda}{4}$, Eq. \eqref{eq:phi_elev} is sufficient to compute the unambiguous elevation angle of a source from a single baseline. For arrays with larger baselines, common interferometric techniques can be used to solve fringe pattern ambiguity using multiple baselines, temporal, or band information~\cite{white_book}.

Inevitably, there is noise in $\Delta \psi$, which affects the computed elevation angle $\phi$. As measurement noise increases, elevation accuracy decreases. The accuracy of $\phi$ is therefore dependent on the \gls{snr} of the signal measurement, which is determined by factors including the brightness of the source, the integration length used in the image formation process, and noise in the imaging system hardware. Source properties cannot be controlled and system properties are fixed, so their effects on \gls{snr} cannot be mitigated.
\gls{snr} is improved with long integration lengths in \gls{sar} measurements, which then improves elevation accuracy. \gls{snr} is further improved by incorporating signal information from multiple \glspl{vx}, using the average phase delay across multiple vertical baselines for $\Delta \psi$ in Eq.~\eqref{eq:phi_elev}. When $N$ vertical baselines are used, \gls{snr} is improved by a factor of $\sqrt{N}$.

\subsection{3D Mapping}
\begin{figure}[!t]
\centering 
\includegraphics[width=0.3\textwidth]{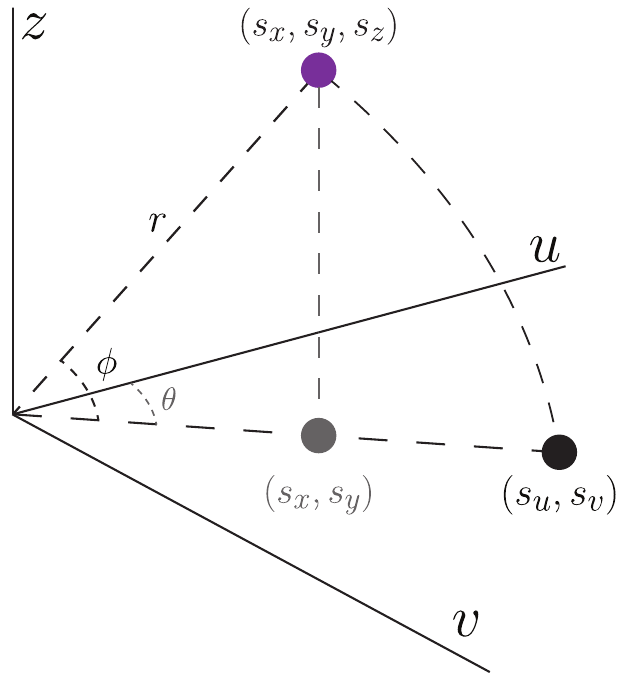}
\caption{Projection Effect: A 3D source at Cartesian coordinates $(s_x, s_y, s_z)$ with range $r$, azimuth $\theta$ and elevation $\phi$ is projected to position $(s_u, s_v)$ in range-azimuth $(u, v)$ space. As elevation increases, the projected position is further out than its 2D Cartesian position $(s_x, s_y)$.}

\label{fig:projection_effect}
\end{figure}

In \gls{sar} processing, there is rotational invariance about the synthetic aperture~\cite{sar_chapter}. This projects elevation along the range dimension. Each 3D source is projected into the \gls{sar} image space, which is a 2D range-azimuth projection space represented by the horizontal $(u, v)$ plane. The source's projected position in range-azimuth space is different from its position in 2D Cartesian space $(x, y)$ as visualized in Figure~\ref{fig:projection_effect}. As elevation angle increases, the difference in position is more prominent, with the projected position of a 3D source further from its 2D Cartesian position. Each source must be de-projected from range-azimuth space in order to correctly combine elevation with range and azimuth.

Combining the range-azimuth signal information from the \gls{sar} image with the elevation information from the interferometric measurement forms a 3D view of the surroundings of a vehicle. This corresponds to a spherical 3D coordinate representation, with range $r$, azimuth $\theta$, and elevation $\phi$. We transform this spherical coordinate representation to \gls{sar}-specific Cartesian coordinates $(s_x, s_y, s_z)$ in order to visualize the scene with a 3D point cloud
\begin{equation}
\label{eq:sphere_cart}
    s_x = r \cos{\theta}, s_y = r \sin{\theta}\cos{\phi}, s_z = r \sin{\theta}\sin{\phi}.
\end{equation}
This process takes advantage of \gls{sar} imaging to increase azimuth resolution. Furthermore, and most importantly, the improved \gls{snr} and source isolation in the \gls{sar} image produces a high-fidelity elevation fit with the elevation extracted with the real array.

\section{Experimental Setup and Processing}
\label{sec:setup}

We apply this approach to agricultural and urban driving scenes, which are primary environments for automotive radar sensors.
Two radar sensor arrays were mounted on a compact car. Each radar was positioned near the front edge of the vehicle and oriented at 45 degrees from the front. This configuration provided 180 degree azimuth coverage across the front of the vehicle. \gls{sar} images were generated independently for both the left and right radar sensors and merged to form images covering the full \gls{fov}. An interferometric processing scheme based on the signal model described in Section~\ref{sec:sigmdl} was used to extract the elevation map and 3D point cloud. Processing time for each image is under one second, with minimal computational cost, making this approach well-suited for future use in real-time automotive \gls{insar}.

\subsection{Radar Array}
\label{sec:radarparameters}
\begin{figure}[!t]
\centering 
\includegraphics[width=0.48\textwidth]{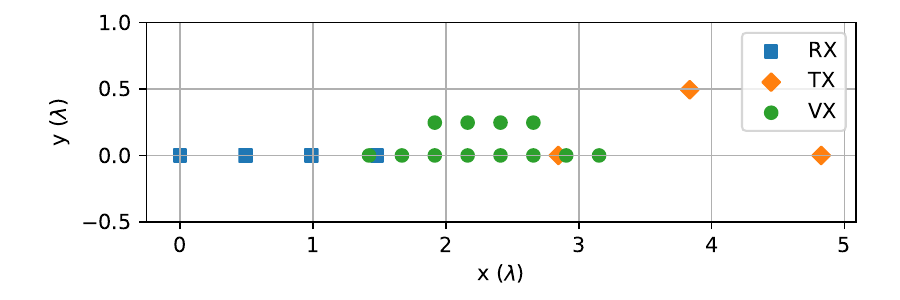}
\caption{Radar Array: TI AWR1243BOOST array consists of 4 \gls{rx} (blue squares) and 3 \gls{tx} (orange diamonds) antenna elements. Antenna elements form a dense, 2-elevation layer virtual array of \glspl{vx} (green circles).}
\label{fig:virtual_array}
\end{figure}

The radar sensor front ends use a TI AWR1243BOOST evaluation board. The AWR1243 operates in the 76--81 \si{\giga\hertz} band and supports 3 \glspl{tx} and 4 \glspl{rx} to generate up to 12 \glspl{vx} in single-sensor mode. The array configuration of the \glspl{tx} and \glspl{rx} on the evaluation board forms a dense virtual array as shown in Figure \ref{fig:virtual_array}. The virtual array has two elevation layers spaced $\lambda/4$ apart. Each three-patch microstrip antenna element provides a wide \gls{fov} in azimuth (78 degrees at half power) and elevation (40 degrees at half power).

\gls{tdm} coding scheme was used to cycle through the 3 \glspl{tx}. \gls{tdm} simplifies processing and maximizes dynamic range but limits the maximum velocity. To use \gls{tdm}, we limit our driving velocity to 20 mph (9 m/s). Extensions to higher driving speeds will be discussed in Section~\ref{sec:conclusion}.

We used a modulation scheme optimized for SAR imaging in urban and agricultural environments. Each chirp was sampled with 512 samples at a rate of 18.75 Msps. The total chirp bandwidth was 820 \si{\mega\hertz} (27.3 \si{\micro\second} pulse length with a 30 \si{\mega\hertz}/\si{\micro\second} ramp slope). The maximum range was 93.7 \si{\meter} with a range resolution of 18.3 \si{\centi\meter}. The \gls{pri} was 63.9 \si{\micro\second} from chirp to chirp, thus for 3 \gls{tx} TDM coding the effective \gls{pri} was 191.7 \si{\micro\second}. Chirp frames were structured to produce 256 chirps per TX with minimal idle time between frames to approximate regular sampling between frames. The modulation center frequency was $f_c=77.4$~\si{\giga\hertz} resulting in a mean wavelength of $\nu_c=3.87$~\si{\milli\meter}, the value used in determining the elevation baseline $D_{\nu}$.

For accurate \gls{sar} processing, we use an off-the-shelf GPS IMU sensor to localize the relative ego-vehicle position to sub-wavelength accuracy~\cite{vn200}. Our acquisition system utilizes a common timestamp using the GPS signal to assign position information to each chirp. Our acquisition system offers two acquisition modes, a raw data recorder and a real-time \gls{sar} imaging system. For this experimental setup we used the raw data recorder, which captures the unprocessed ADC samples, timestamps, and GPS IMU information to file to be processed offline. We will discuss performing this approach in real time in Section \ref{sec:conclusion}, which would be possible with little additional computational cost.

\subsection{SAR Imaging}
\gls{sar} images are formed per \gls{vx} on the real array using the \gls{fbp} algorithm~\cite{sar_chapter}. We use standard approximations in SAR imaging, and make the start-stop approximation across the complete TDM cycle. We image over a 1~m aperture, empirically chosen to balance maximizing the aperture length while minimizing the GPS IMU errors. For each \gls{vx}, a 30 \si{\meter} x 30 \si{\meter} image is formed with a pixel resolution of 4 \si{\centi\meter} to oversample the range resolution. Each pixel is complex-valued, maintaining crucial phase information for the interferometric measurement.

\gls{snr} is improved by incorporating signal information from multiple \glspl{vx} simultaneously in \gls{fbp}. Furthermore, interferometric measurements from multiple vertical baselines are averaged to increase \gls{snr}. This utilizes the maximum number of \glspl{vx} possible in our radar array (Figure~\ref{fig:virtual_array}).

\subsection{Elevation Mapping and Processing}
\label{sec:elevationmappingprocess}

After SAR image formation, elevation is extracted from the interferometric measurement by applying Equation~\eqref{eq:phi_elev}. Elevation is then combined with the range and azimuth information at each \gls{sar} image pixel to form a 3D representation of the scene. This initial cluster of points has no distinction between signal from real sources or noise. Noise and spurious effects are inevitable since data is acquired in real automotive environments. To extract scene structure, we use a custom series of signal processing operations, highlighting high-\gls{snr} information from strong sources, corresponding to real features, while suppressing low-\gls{snr} information from noise.

An \gls{snr} threshold of 15 \si{\decibel} separates real sources from noise. We compute \gls{snr} by comparing per-pixel signal magnitude to the median signal across the scene, as the median is more robust to outliers than the mean for large data. Since multiple baselines were averaged to improve \gls{snr}, points are further filtered based on phase variations across the \glspl{vx} in these baselines. Removing these high variation points suppressed false arcing structures from strong point responses. 

The radar has high sensitivity to close-range objects. Elevation angles exceeding 45 degrees were removed, based on the largest angles for typical scene objects and the radar \gls{fov}. The close-range sensitivity resulted in a constant cluster of points appearing in front of the car. This cluster was removed with a minimum radius cutoff of 2 meters and an azimuth cutoff of $\pm$ 15 degrees from the front. Due to multiple path effects and reflections from the ground, some points accumulated phase corresponding to points underground, which cannot be detected in our environment and with our experimental setup. A minimum elevation threshold was used to remove points that appeared underground. 

After signal post-processing, the spherical coordinate representation is transformed to Cartesian coordinates using Equation~\eqref{eq:sphere_cart}. The resulting points in the scene are stored in a point cloud~\cite{pypcd}. Point clouds are visualized using PCL Tools~\cite{PCLlib}.

\subsection{Accuracy Test}

To validate our elevation mapping approach in a controlled environment, we set up a simple scene in a test chamber to study the accuracy. We placed one radar array on a tabletop rail system, mounted perpendicular to the direction of travel and formed a 1~m aperture. All other acquisition parameters are as described in Section~\ref{sec:radarparameters}.

\begin{figure}[!t]
\centering 
\includegraphics[width=0.5\textwidth]{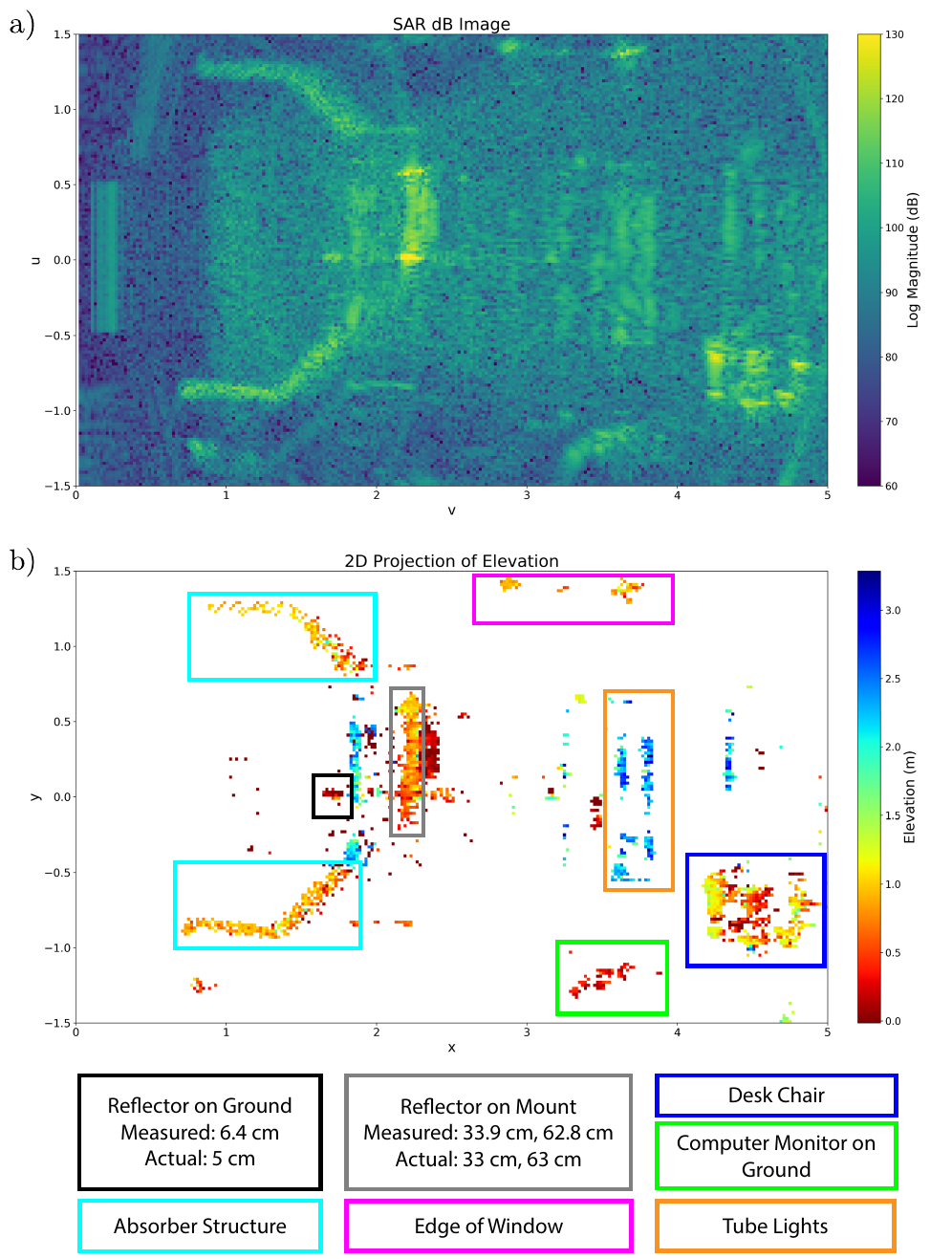}
\caption{Reflector Accuracy: a) \gls{sar} log magnitude image of test scene with two reflectors. b) 2D projection map with points colored by elevation in meters. One reflector is placed on the ground and the other is placed on a mount at either 33 \si{\centi\meter} or 63 \si{\centi\meter}. This projection map corresponds to the 63 \si{\centi\meter} case. Other objects in the scene include an absorber structure around the reflectors, tube lights, a desk chair, a window, and a computer monitor, each labeled with a colored box.}

\label{fig:cnc_room}
\end{figure}

This simple scene contained two high \gls{rcs} corner reflectors. One reflector was placed on the ground and the other reflector was placed behind it on a tripod mount with an extended arm and adjustable height. We tested two height positions for the reflector mounted on the tripod: 33 \si{\centi\meter} and 63 \si{\centi\meter}. The \gls{sar} image corresponding to this scene with the mounted reflector at 63 \si{\centi\meter} is shown in Figure~\ref{fig:cnc_room}a. An absorber structure surrounds the two reflectors. The scene also contains overhead tube lights, a desk chair, a computer monitor, and a window.

Figure~\ref{fig:cnc_room}b shows the 2D Cartesian projection of the scene, where points are colored by elevation in meters. 0 \si{\meter} corresponds to ground elevation. We can see the objects in the room are well localized in elevation. Each object in the room is outlined with a colored box and a corresponding label. The mounted reflector is outlined with a gray box. When the mounted reflector was placed at 33 \si{\centi\meter}, the measured elevation with our elevation mapping approach was 33.9 \si{\centi\meter}. When the mounted reflector was placed at 63 \si{\centi\meter}, the measured elevation with our elevation mapping approach was 62.8 \si{\centi\meter}. The reflector on the ground is outlined with a black box. The center of this reflector is 5 \si{\centi\meter} off of the ground, and the measured elevation with our elevation mapping approach was 6.4 \si{\centi\meter}. The extended objects in the room also have sensible elevations. The tube lights are above 2 \si{\meter}, the desk chair is up to 1 \si{\meter}, the computer monitor on the ground is below 0.5 \si{\meter}, the window edge is at around 1 \si{\meter}, and the absorber structure is at maximum brightness at the table height as that is the antenna boresight.

\section{Results and analysis}
\label{sec:results}

We apply this approach to agricultural and urban driving scenes. For the agricultural use case, we demonstrate driving through a vineyard on a poorly-paved road. For the urban use case, we show a typical driving scene: an intersection in a dense environment with other vehicles, buildings, and trees.
For each use case, we display a camera view of the scene and a high-resolution \gls{sar} image. We generate a 2D Cartesian projection of the 3D scene representation by using the elevation mapping procedure outlined in Section~\ref{sec:setup} and projecting across $z$. In the 2D Cartesian projection, an elevation of 0 \si{\meter} corresponds to the mounted position of the radar arrays on the vehicle, which is slightly above ground.

\subsection{Agricultural Driving}

\begin{figure*}[!t]
\centering 
\includegraphics[width=0.87\textwidth]{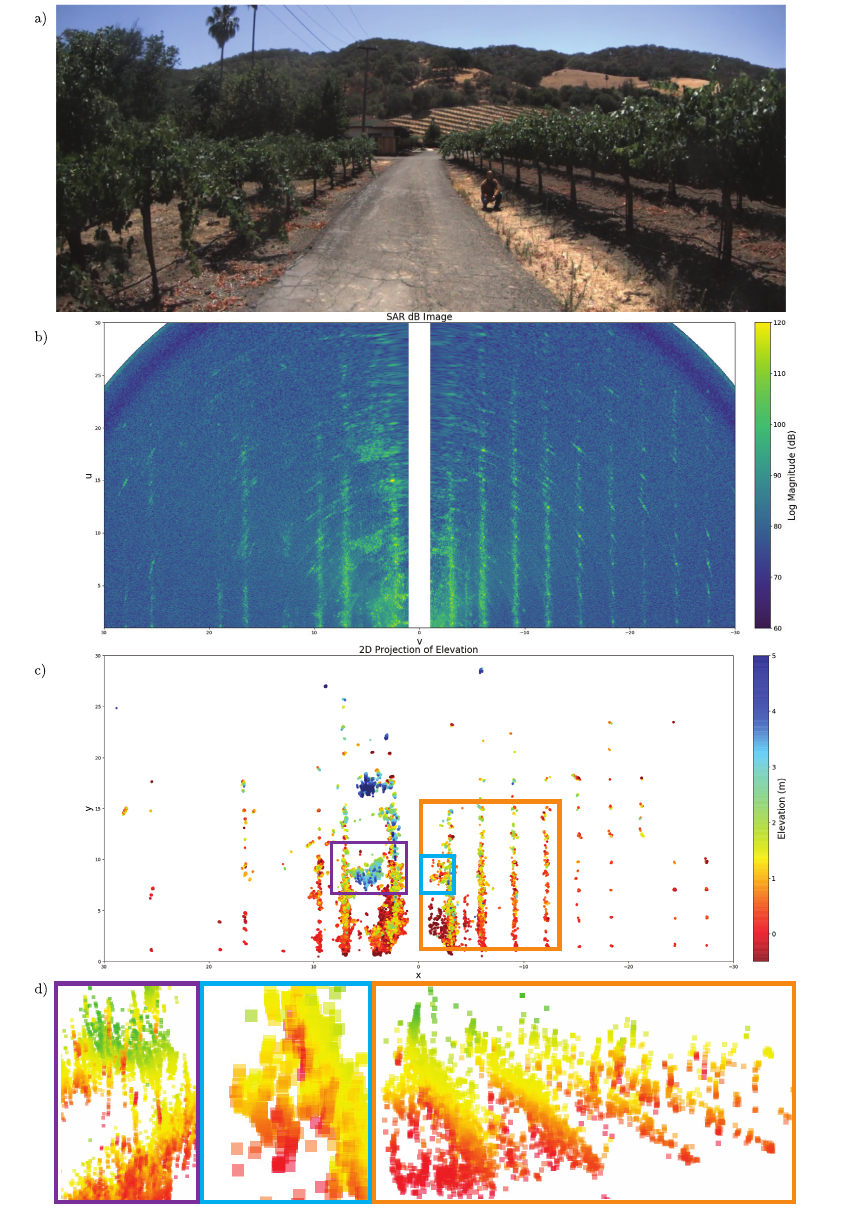}
\caption{Vineyard Scene: a) Camera view containing features of interest including rows of grapevines, trees, and a person. b) SAR log magnitude image. c) 2D projection map with points colored by elevation in meters and boxes around regions of interest. d) Point cloud renderings of regions of interest. Points colored by elevation according to the color scale in c). The left purple box contains a cluster of trees with visible tree trunks, appearing above the rows of grapevines. The middle blue box contains a person kneeling down next to the grapevines. The right orange box contains rows of grapevines.}
\label{fig:vineyard}
\end{figure*}

The camera view of the agricultural scene is shown in Figure~\ref{fig:vineyard}a. There are rows of grapevines on both sides of the road, a person crouching next to the road on the right, and clusters of taller trees among the grapevines on the left. Figure~\ref{fig:vineyard}b shows the high-resolution log-magnitude \gls{sar} image. The rows of grapevines are bright vertical lines and trees are round clusters, both in green. Both trees and grapevines cause strong point response and projection effects, producing arcing structures.  

Figure~\ref{fig:vineyard}c contains the 2D Cartesian projection image, where points are colored by elevation in meters. The grapevines have consistent elevation of approximately 1.5 \si{\meter}, with clusters of trees represented by blue and green points at higher elevation than the grapevines. The point response and projection artifacts that appear in the \gls{sar} image are corrected. Points at higher elevation are more concentrated and arcing artifacts on points in the grapevines are reduced.

Figure~\ref{fig:vineyard}d highlights insets from the 3D point cloud of this scene. The purple box on the left contains the cluster of trees from the left corner in the camera view in Figure~\ref{fig:vineyard}a. The grapevines surrounding the trees are at consistent elevation, represented by red and orange points. The tree trunks appear as thin columns in the center, colored by red and orange, with branches outlined by yellow and green points that reach upwards. The blue box in the middle isolates the person crouching next to the grapevines. The dark red points clustered in the center correspond to the person, with yellow points for the grapevines on the right and the grapevine branch reaching out behind the person. The orange box on the right shows the rows of grapevines on the right side of the road. Each row of grapevines has consistent elevation, with yellow points at the top. At the ends of the rows there are scattered green points from stray grapevine branches, and the irrigation systems and weeds on the ground are represented with dark red clusters.

In this agricultural scene we demonstrate successful and consistent extraction of elevation. The long-range capabilities of \gls{sar} are highlighted by the many vineyard rows with consistent elevation, even though the number of high-\gls{snr} points decrease further from the sensors. The elevations of features in the scene, such as grapevines, are as expected. Trees are clearly separated from the vineyard rows below. Small details that are close together, such as the person in front of the grapevine, are distinguishable but push the limits of the single source per range-azimuth bin approximation.

\subsection{Urban Driving}

\begin{figure*}[!t]
\centering 
\includegraphics[width=0.87\textwidth]{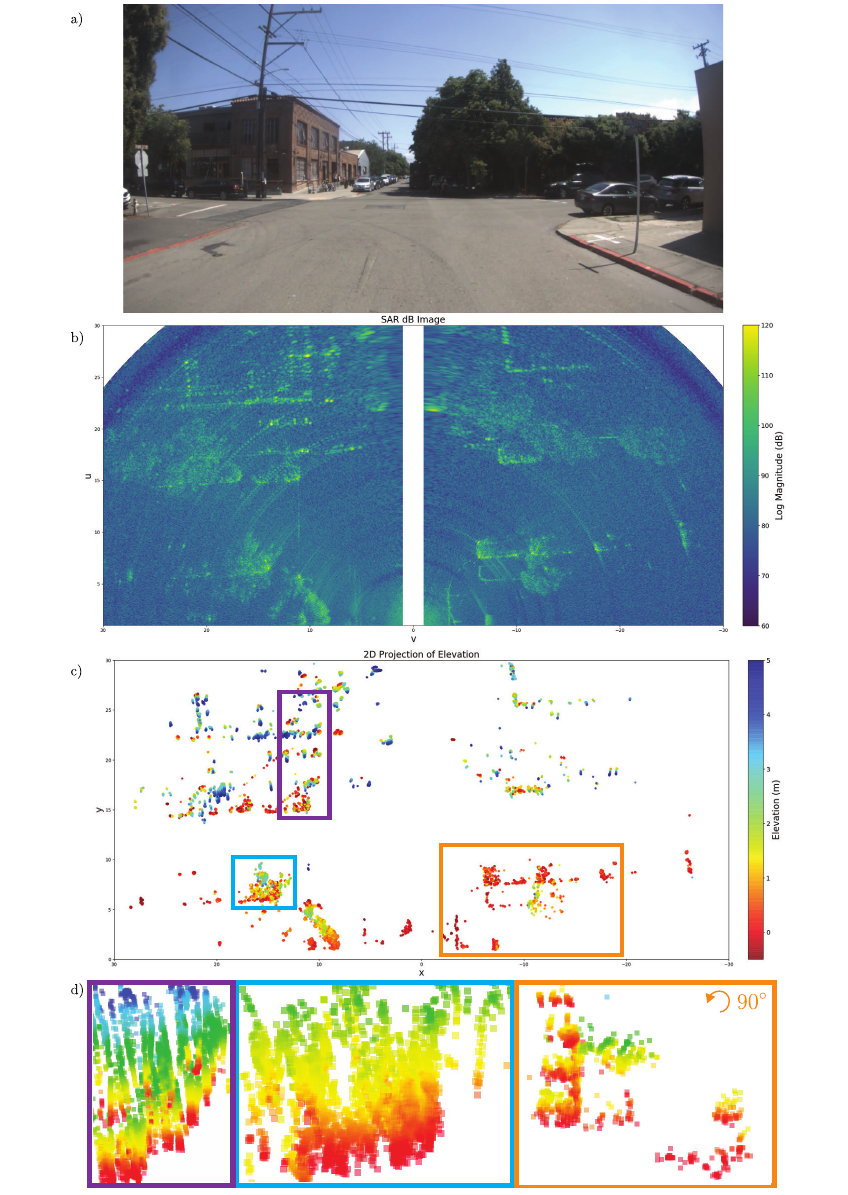}
\caption{Urban Scene: a) Camera view containing features of interest including cars, trees, and buildings. b) SAR log magnitude image. c) 2D projection map with points colored by elevation in meters and boxes around regions of interest. d) Point cloud renderings of regions of interest. Points colored by elevation according to the color scale in c). The left purple box contains a car, electrical post, and a building section with regularly-spaced pillars. The middle blue box contains a car and tree. The right orange box, rotated 90 degrees counterclockwise relative to the projection map, contains a curb, pole, tree, and three cars.}
\label{fig:urban}
\end{figure*}

The camera view of the urban scene is shown in Figure~\ref{fig:urban}a. There is a four-way intersection with parked cars, trees, poles, curbs, and a tall building with pillars and grids of windows. Figure~\ref{fig:urban}b shows the high-resolution log-magnitude \gls{sar} image. The outlines of parked cars are visible. Thin lines denote sidewalks and curbs, sharp right-angled structures are building features, and diffuse clusters are trees.

Figure~\ref{fig:urban}c displays the 2D Cartesian projection image, in which points are colored by elevation in meters. Cars have consistent elevations around 1 \si{\meter}, tree branches appear starting at 2 \si{\meter}, and the two-story building is around 5 \si{\meter}, which are all reasonable elevations for the objects. With the \gls{snr} threshold, there are fewer points captured for fine features such as sidewalks and curbs, but those captured points are correctly located on the ground.

In Figure~\ref{fig:urban}d, insets from the urban scene are highlighted. The purple box on the left displays the side of the two-story brown building visible on the left side of Figure~\ref{fig:urban}a. The grid of pillars and windows are represented by the vertical columns that reach from the ground (in red) to the top of the building (in dark blue). The protrusion of each pillar in front of the windows is captured. In the front of this inset is a red and yellow cluster representing the end of the car and the telephone pole next to it. Red points on the side correspond to the ramp and bike next to the building. The blue box in the middle displays a zoom in on the white car and stop sign on the left side of the intersection, with a tree reaching above the car. These sources are not completely visible in the camera view in Figure~\ref{fig:urban}a, but are clearly captured in the \gls{sar} image in Figure~\ref{fig:urban}b. The car is the rounded red and yellow cluster. The yellow and green points are the large tree branch reaching over the car and out towards the road, with a branch bending down to the right of the car. The orange box on the right contains three cars, a curb, a pole, and a tree. This inset is rotated 90 degrees counterclockwise relative to the camera view for better point cloud visualization. The three red and yellow clusters corresponding to cars have consistent elevation. The detailed outline of each car is correctly captured, with the roof in yellow and the hood in red. Next to the first car, a right-angled structure of dark red points at ground level corresponds to a garden bed with weeds on the sidewalk. Next to the middle car, there is a tree in the garden bed, represented by the cluster of yellow and green points, at higher elevation than the cars. Sections of the curb next to the main road are captured in the bottom left of the inset, with a thin yellow line corresponding to the pole, and the yellow cluster corresponding to the corner of the tan-colored building.

This scene shows isolation of sources in a detailed, large-range scene with lots of information. There are many objects and structures that result in complex interactions, reducing the number of high-\gls{snr} signal points extracted. With fewer points, some details such as small curbs are not fully outlined. However, there are more complicated features available in this urban scene than in the agricultural scene, and many objects and details are extracted clearly, such as the cars and tree in the right inset in orange in Figure~\ref{fig:urban}d. Separation of objects that are close together, such as the car and tree above in the middle inset in blue in Figure~\ref{fig:urban}d, is limited due to the single source per range-azimuth bin approximation. We successfully produced detailed elevation maps of a car's drivable region for multiple use cases.

\section{Conclusion}
\label{sec:conclusion}

We have shown that Interferometric \gls{sar} processing can be used to enhance the elevation accuracy and \gls{snr} of off-the-shelf radar arrays used in common human-scale autonomous vehicle environments such as agricultural processing and urban driving. We demonstrated high-quality automotive elevation mapping with \gls{insar}, forming 3D point cloud representations of urban and agricultural environments using a simple, interpretable, and low computational cost processing approach. This is a new application of \gls{insar}, accounting for the complex trajectories and vehicle dynamics that are associated with driving in diverse automotive environments.

This approach can easily be incorporated in a real-time \gls{sar} imaging system \cite{sar_chapter}.
In this work, we processed captured data offline. Applying our elevation mapping procedure took less than one second per \gls{sar} image frame. Optimizing the computation of each phase extraction and signal filtering operation will provide further improvements to computation speed.
Using this approach in real time will provide a low-latency point cloud of static objects. Static object point clouds, combining elevation information with enhanced azimuth resolution from \gls{sar}, would be an excellent starting point to providing a drivable area occupancy grid for perception and inference in autonomous driving \cite{zendardataset, 12145}. The point cloud produced in our approach is primarily of stationary objects because moving objects are well-filtered in \gls{sar} images. Traditional range-Doppler processing~\cite{range-doppler} would complement our \gls{sar} processing approach by providing detection and localization of moving objects. 

The low elevation resolution in our array is a dominant limiting factor for our elevation accuracy. In our controlled environment studies, we were able to localize nearby objects to centimeter-level accuracy even with the low elevation resolution. This is because the accuracy is related to the aperture size and the S/N of the object. In the on-vehicle tests, the range of objects was extended leading to lower S/N and lower accuracy. We used a manufacturer-supplied radar array optimized for simplicity and azimuth resolution. This is not the ideal array for our approach, as the azimuth resolution is dominated by the synthetic aperture response. A more optimal array design would primarily focus on elevation resolution. For the most flexibility in use, the ideal array would still maintain some azimuth resolution to enable moving object tracking with traditional range-Doppler processing. 

With a higher elevation resolution array, we would expect vibrations of the array (e.g. due to rough roads) to become a limiting factor. Elevation accuracy would be improved by modeling vibration patterns and using an array geometry with a larger physical elevation aperture.

For our experimental set up we used \gls{tdm} as a simple form of MIMO coding. This limits the maximum velocity of the vehicle before aliasing artifacts start appearing in the SAR image. A natural extension of this work is to use a coded MIMO scheme. This would have the benefit of extending the maximum velocity and potentially increasing sensitivity over a fixed aperture. 

We used a simple phase difference model to determine the elevation angles for each \gls{sar} image pixel. This model used only the vertical baselines in an array. Using baselines with various orientations in a generalized approach could further enhance elevation resolution and sensitivity. Furthermore, our model made the approximation of a single source per range-azimuth bin. This approximation held for most portions of the agricultural and urban environments, but had limitations for objects that were close together, such as the person next to the grapevine in Figure~\ref{fig:vineyard}d and the car and tree in Figure~\ref{fig:urban}d. A more advanced model, such as maximum likelihood estimation or iterative deconvolution \cite{Cornwell_2008}, could be used in order to fit and resolve multiple sources per range-azimuth bin. 

\bibliographystyle{IEEEtran}
\bibliography{references}

\end{document}